\documentclass[aps,floatfix,twocolumn,footinbib,superscriptaddress,showpacs]{revtex4}
\usepackage{epsfig}
\usepackage{amsmath,amssymb}
\usepackage{bm}
\begin{document}

\title{Bias-induced destruction of ferromagnetism and disorder effects in GaMnAs heterostructures}


\author{Christian Ertler\footnote{email:christian.ertler@uni-graz.at}}

\author{Walter P\"otz}
\affiliation{Institute of Theoretical Physics, Karl-Franzens University Graz,
Universit\"atsplatz 5, 8010 Graz, Austria}

\pacs{85.75.Mm, 73.23.Ad, 73.63.-b, 72.25.Dc}

\begin{abstract}

The magneto--electric properties of resonant tunneling double barrier structures using  GaMnAs for 
the quantum well is investigated within a self-consistent Green's function approach and  a tight--binding electronic structure model. The magnetic
state of the well is determined self--consistently by the tunneling current which controls the hole spin density and, hence, the degree of
exchange splitting of the subbands inside the well. 
Prompted by recent experiments we compare model systems of increasing defect concentration (substitutional disorder) regarding their I-V curve, magnetic state, and spin polarization.
We predict that, near resonance, the ferromagnetic order which may be present at zero bias  in the GaMnAs well tends to be destroyed.  
Resonance peaks are found to be more sensitive to disorder than ferromagnetic ordering and spin polarization of the steady--state current.

\end{abstract}

\maketitle

\section{Introduction}
\label{intro}

The realization of electric control of ferromagnetism in
nanostructures is of great interest both for spintronic device
application and for achieving a better understanding of the
physical mechanisms and dynamics underlying the formation of ferromagnetic order in Mn doped semiconductors. 
Dilute magnetic semiconductors (DMS) are made magnetic by doping of
ZnS--structured  semiconductors with transition metal
elements, which provide local magnetic moments arising from open electronic $d$ or $f$ shells \cite{Jungwirth2006:RMP,Burch2008:JMMM}.
A prototype is bulk Ga$_{1-x}$Mn$_x$As in which Mn residing on the Ga site (Mn$_{\mathrm{Ga}}$) donates  both a hole and a local magnetic moment.    Mn$_{\mathrm{Ga}}$ is an at least moderately deep acceptor and associated levels lie
about 100 meV above the valence band edge \cite{Schneider1987:PRL}.   
 In a recent scanning tunneling
microscopy experiment the radius of the Mn acceptor wave function
has been determined  to be about 2~nm \cite{Richardella2010:S}. 
Due to an antiferromagnetic exchange coupling
between the itinerant holes and the local Mn $d$-electrons, an effective ferromagnetic ordering among the Mn-ions, known as carrier--mediated ferromagnetism,  can be established \cite{Ohno1996:APL,VanEsch1997:PRB,Dietl2000:Science}.
Due to the hole--concentration--dependent effective exchange field the spin degeneracy of the holes is
lifted resulting in a self--consistently spin polarized hole gas.  The critical temperature $T_c$ for the occurrence of ferromagnetism in bulk GaMnAs is typically below $\sim 150$ K and depends on Mn concentration and sample preparation \cite{Jungwirth2006:RMP,Burch2008:JMMM,VanEsch1997:PRB}.

Probably due to the degree of (unwanted) defects in GaMnAs samples depending on growth
conditions, experimental evidence has led to 
somewhat conflicting conclusions about the electronic structure in the vicinity of the Fermi level. While all experimental studies on ferromagnetic bulk GaMnAs confirm it to be p--type, there is some debate as to the precise position of the Fermi energy \cite{Burch2008:JMMM}. Some experiments can be interpreted by placing it into the top of a GaAs--like valence band edge which, at most,  broadened by disorder \cite{Jungwirth2006:RMP}. Others suggest
the existence of an isolated impurity band (of high ``effective mass'') which forms at Mn concentrations
above $\sim 1.5\%$ leading to a metal--insulator transition in
high--quality GaMnAs \cite{Richardella2010:S,Burch2006:PRL,Ohya2011:NP}.

Irrespective of the detailed electronic structure it is clear that, due to disorder, the localization length of valence band eigenstates will tend to decrease as one moves from the top of the valence band towards the energy gap \cite{Madelung:1978}. 
Near the band edge a coexistence of localized and delocalized (Bloch--like) eigenstates may be expected, similar to amorphous Si \cite{Miyazaki1987:PRL,Li1993:PRB}.    If Mn$_{\mathrm{Ga}}$ is the main defect to provide modification of the valence band edge from what it is for GaAs, 
quantization effects can be expected for a layer thickness $\leq$ 3 nm.   Recent
tunneling spectroscopy of GaMnAs quantum well structures has
indicated such effects \cite{Ohya2011:NP}. However, the signatures
in the current--voltage characteristics appear to be rather weak and
no regions of negative differential conductivity due to resonances
associated with GaMnAs well layers have been observed yet, with the exception of an asymmetric magnetoresistance resonant
tunneling structure \cite{Likovich2009:PRB}. This
suggests that a significant concentration of defects may be present, as
it is the case, e.g., in thin layers of amorphous Si, in which
similarly weak signatures have been
found \cite{Miyazaki1987:PRL,Li1993:PRB}. 
Disorder on the other hand, as we know from bulk  Ga$_{1-x}$Mn$_x$As, goes hand in hand with ferromagnetic order and, indeed, ferromagnetic behavior has been verified experimentally for thin layers of GaMnAs \cite{Ohya2011:NP,Mathieu2003:PRB}.

Due to
spin--selective hole tunneling in and out of a {\it ferromagnetic} Mn-doped GaAs
quantum well, regardless whether occurring sequentially or resonantly,  one can expect hole population and spin polarization to become dependent upon the 
bias applied to a double barrier structure containing such a well.  This mechanism 
should allow an electric control of ferromagnetic order in the quantum well and is evaluated in this paper with particular attention to the qualitative effect of disorder in the Mn doped well region.   An important question addressed is whether disorder and unwanted defects can suppress {\it spin--selective tunneling}.   Disorder has a strong influence on the electronic properties and they, in turn, strongly influence (resonant) tunneling.  
High--quality ferromagnetic layers integrated in semiconductor
heterostructures,  have been envisioned to allow
spin-dependent carrier transmission \cite{Sankowski2007:PRB}. Magnetic resonant tunneling structures of
high structural quality may
 allow the realization of spin
valves, spin filtering, and spin switching devices, as proposed in
several studies
\cite{Likovich2009:PRB,Slobodskyy2003:PRL,Slobodskyy2007:APL,Ertler2006a:APL,Ertler2007a:PRB,Petukhov2002:PRL,Ohya2010:PRL,Ohya2010:APL}.

In this article we investigate spin--selective hole transport in
GaAs/AlGaAs/GaMnAs/AlGaAs/GaAs hetero-structures within the limit of moderately thin samples
so that an effective independent particle model provides a good first approximation.  
We apply a
non-equilibrium Green's function formalism based on a tight--binding
Hamiltonian for the electronic structure, including self-consistency
regarding the charge density and the exchange splitting of the
effective potential, as well as charge transfer to the contacts.
The carriers' Coulomb interaction and the exchange coupling
with the magnetic ions are described within a mean-field picture.
Details of our model are exposed in Sect.~\ref{sec:model}. 
Since disorder seems to play a major role in actual samples
we study the effect of substitutional  disorder on the I-V characteristics, the ferromagnetic state of the heterostructure, and spin--polarization 
of the current density.  
Results and relevance to
experiment are discussed in
Sect.~\ref{sec:results}.
Summary and conclusions are given in Sect.~\ref{sec:sum}.

\section{Selfconsistent Transport  Model}
\label{sec:model}

The basic features of the semiconductor double--barrier structure  near the top of the valence band edge are mapped onto a
two--band tight-binding Hamiltonian for the heavy holes $(J_3=\pm
3/2)$
\begin{eqnarray}
H_s &=& \sum_{i,\sigma} \varepsilon_{i,\sigma} |i,\sigma\rangle\langle
i,\sigma|\nonumber\\
&&+\sum_{i,\sigma\sigma'}t_{i,\sigma\sigma'}|i,\sigma\rangle\langle
i+1,\sigma'|+ \mathrm{h.c.},
\end{eqnarray}
where  $\varepsilon_{i,\sigma}$ is the spin-dependent ($\sigma
=\uparrow,\downarrow \equiv \pm 1$) onsite energy at lattice site
$i$, $t_{i,\sigma\sigma'}$ denotes the hopping-matrix between
neighboring lattice sites, and $\mathrm{h.c.}$ abbreviates the
Hermitean conjugate term. Spin conserving hopping gives a diagonal
matrix $t_{i,\sigma,\sigma'} = t\delta_{\sigma\sigma'}$ with the
hopping parameter $t = -\hbar^2/(2 m^* a^2)$ depending on the
effective mass $m^*$ and the lattice spacing $a$ between to
neighboring lattice sites. The onsite energy
\begin{equation}
\varepsilon_{i,\sigma} = U_i - e \phi-\frac{\sigma}{2}\Delta_i
\end{equation}
includes the intrinsic hole band profile $U_i$ due to the band
offset between different materials, the electrostatic potential
$\phi$ with $e$ denoting the elementary charge, and the local
exchange splitting $\Delta_i$.  Near the band-edges this model is
equivalent to an effective--mass model, however, it has the
advantage that structural imperfections, as well as spin--flip processes, can be readily be modeled  by
varying the onsite and hopping energies. A more realistic 
description can be achieved by introducing a larger set of orbital
basis functions at each lattice site
\cite{Schulman1983:PRB,DiCarlo1994:PRB,Poetz1989:SM,Sankowski2007:PRB}.

Within a mean-field approach the exchange coupling between holes and magnetic
impurities can be described by two interrelated effective
magnetic fields,
respectively, originating from  a nonvanishing mean spin
polarization of the ions' d--electrons  $\langle S_z\rangle$ and from the hole
spin density $\langle s_z\rangle = (n_\uparrow-n_ \downarrow)/2$ \cite{Dietl1997:PRB,Jungwirth1999:PRB,Fabian2007:APS}.
The exchange splitting of the hole bands is  given
by
\begin{equation}\label{eq:delta}
 \Delta(z) = -J_\mathrm{pd} n_\mathrm{imp}(z) \langle S_z\rangle(z)~,
\end{equation}
with $z$ denoting the longitudinal (growth) direction of the structure,
$J_\mathrm{pd} > 0 $  is the exchange coupling  between the impurity
spin and the carrier spin density (in case of GaMnAs p-like holes
couple to the d-like impurity electrons), and $n_\mathrm{imp}(z)$ is
the impurity density profile of magnetically active ions. The
magnetic order between the impurities is mediated by the holes and the
effective impurity spin polarization depends on the mean hole spin
polarization via
\begin{equation}\label{eq:Szgen}
  \langle S_z\rangle= - S B_S\left( \frac{S J_\mathrm{pd} \langle s_z \rangle}{k_B T}\right),
\end{equation}
where, respectively,  $k_B$,  $T$,  $B_S$ is the  Boltzmann constant, the lattice
temperature, and  the Brillouin function of order $S$, here with
$S = 5/2 $ for the Mn impurity spin. Combation of  Eq.~(\ref{eq:delta}) and
Eq.~(\ref{eq:Szgen}) gives a self-consistent effective Hamiltonian
for the holes $H_\mathrm{eff} = -\sigma \Delta(z)/2$ with
\begin{equation}\label{eq:delta1}
 \Delta(z) = J_\mathrm{pd} n_\mathrm{imp}(z)
S B_S\left\{\frac{S J_\mathrm{pd} [n_\uparrow(z)-
n_\downarrow(z)]}{2 k_B T}\right\}.
\end{equation}
This shows that a manipulation of the hole spin density $\langle s_z\rangle$ by the applied bias is the key to the control  of ferromagnetic order in the heterostructure.

Within a Hartree mean-field picture space-charge effects are taken into account
self-consistently  by calculating the electric potential from the Poisson equation,
\begin{equation}\label{eq:poisson}
 \frac{\mathrm{d}}{\mathrm{d}z} \epsilon \frac{\mathrm{d}}{\mathrm{d}z}\phi =
e\left[ N_a(z) - n(z)\right],
\end{equation}
where $\epsilon$ and $N_a$, respectively, denote the dielectric constant and the
Mn$_{\mathrm{Ga}}$  density. The
local hole density at site $|i\rangle$ is computed as
\begin{equation}\label{eq:n}
 n(i) = \frac{-i}{A a}\sum_{k_{||},\sigma}\int\frac{\mathrm{d}E}{2\pi} G^<(E;i\sigma,i\sigma)~,
\end{equation}
with $A$  and $k_{||}$, respectively, being the in-plane cross sectional area of the structure and the in-plane momentum. 
The non-equilibrium ``lesser''
Green's function $G^<$ is calculated from the equation of motion
\begin{equation}\label{eq:gless}
 G^< = G^R\Sigma^<G^A
\end{equation}
where $G^R$ and $G^A = [G^R]^+$ denotes the retarded and advanced Green's function, respectively.
The scattering function $\Sigma^<=\Sigma^<_l+\Sigma^<_r$ describes
particle inflow of from the left $(l)$ and right $(r)$ reservoir and  \cite{Datta:1995}
\begin{equation}
\Sigma^<_{l,r} = f_0(E-\mu_{l,r})(\Sigma^A_{l,r}-\Sigma^R_{l,r})~,
\end{equation}
where $f_0(x) = [1+\exp(x/k_B T)]^{-1}$ is the Fermi distribution function and
$\mu_l$ and $\mu_r$, respectively, denote the quasi--Fermi energies in the contacts.
The retarded and advanced self-energy terms $\Sigma^R = \Sigma_l^R+\Sigma_r^R$ and $\Sigma^A = [\Sigma^R]^+$, respectively, 
couple of the simulated system region to the left and right contacts, modeled as semi--infinite chains, for which 
analytic expressions are available \cite{Datta:1995,Economou:1983}.
The retarded Green's function, finally,  is given by
\begin{equation}\label{eq:gr}
 G^R = \left[E+i\eta-H_s-\Sigma^R\right]^{-1}~.
\end{equation}

A self--consistent calculation of the spin--dependent effective (one--particle) potential, 
based on the band splitting given by
Eq.~(\ref{eq:delta1}),
 the Poisson equation Eq.~(\ref{eq:poisson},\ref{eq:n}),
and the kinetic equations Eqs.~(\ref{eq:gless}) and (\ref{eq:gr}), 
also  entails 
an adjustment of the quasi--Fermi energies relative to the band
edges in the contacts  to ensure asymptotic charge
neutrality \cite{Poetz1989:JAP}.    These loops must be solved self-consistently until convergence to a
steady--state solution is reached.  A small external magnetic field is applied initially to aid spontaneous symmetry breaking.  
For the next bias iteration, the self--consistent solution from the previous bias value is used for an initial guess.  In most cases a stable self--consistent 
solution is found.  However, due to the nonrelativistic nature of the model with orbital motion decoupled from the spin degree of freedom, "spin up" and "spin down"  are degenerate.  Moreover, 
there are bias regions where no single stable solution exists regarding charge distribution (intrinsic charge bistability) and/or  magnitude of the exchange interaction (magnetic multi-stability) \cite{Poetz1989:JAP,Ertler2010:APL}.

Having obtained the self-consistent potential profile
the spin-dependent transmission probability $T_{\sigma'\sigma}(E)$
from the left to the right reservoir is calculated
from special matrix elements of the retarded Green's function \cite{DiCarlo1994:PRB}
\begin{equation}
T_{\sigma'\sigma}= T_{\sigma'\leftarrow\sigma}(E) =
\frac{v_{r,\sigma'}|G^R(E;r\sigma',l\sigma)|^2}{v_{l,\sigma}
|G^0(E;l\sigma,l\sigma)|^2}
\end{equation}
with $G_0$ denoting the free Green's function of the asymptotic
region, and $v_{l,\sigma}$ and $v_{r,\sigma}$, respectively, are the
spin-dependent group velocities in the leads.
$G^R(E;r\sigma',l\sigma)$ is computed most conveniently by adding
one layer after the other which requires solely 2x2 matrix inversions
within  the present two--band model \cite{Economou:1983}.

The steady--state current
is calculated within stationary  scattering theory (a generalized Tsu-Esaki formula),
\begin{eqnarray}
 j_{\sigma'\sigma} & = & \frac{e m^* k_B T}{(2\pi)^2\hbar^3}
\int_0^\infty \mathrm{d} E\: T_{\sigma'\sigma } g(E)\nonumber\\
g(E) & = & \ln\left\{\frac{ 1 + \exp\left[(\mu_l-E)/k_B T\right]}{
1 + \exp\left[(\mu_r-E)/k_B T\right]}\right\}.
\end{eqnarray}
The applied bias $V=(\mu_l-\mu_r)/e$ is defined as the difference
in quasi-Fermi levels of the contacts. In such a nonequlibrium
situation the hole concentration in the GaMnAs layer is not  determined not simply by its ferromagnetic state, but also depends on other 
physical quantities, such as doping level in the contact regions, structural parameters of the heterostructure, and applied bias.

\section{Results and Discussion}\label{sec:results}

In this section we employ the model outlined above and study the influence of disorder on the resonant tunneling characteristics and the magnetic
state of a double-barrier structure featuring a GaMnAs quantum well.
For this simulation we use  parameters for GaAs and GaMnAs which are we well established in the literature:
 $m^* = 0.4\:m_0$, $\epsilon_r = 12.9$, $V_\mathrm{bar} = 300$ meV, $\mu_l =\mu_r= 70$ meV, $d = 20 \AA$,
$w = 30 \AA$, $n_\mathrm{imp} = 1\times10^{20} $cm$^{-3}$,
$J_{\mathrm{pd}} = 0.15$ eV nm $^3$ \cite{Lee2000:PRB}, $T = 4.2$~K,
where $m_0$ denotes the free electron mass,  $\epsilon_r$ is the
relative permittivity, $V_\mathrm{bar}$ is the bare barrier height
of AlGaAs relative to GaAs, $d$ and $w$, respectively, are the
barrier and quantum well width. The background charge $N_a$ arises from 
an assumed  10\% of the Mn doping $n_\mathrm{imp}$ , since
GaMnAs is known to be a heavily compensated semiconductor\cite{VanEsch1997:PRB,DasSarma2003:PRB}. 
Doping in the contacts, experimentally Be may be used, is such that the thermal
equilibrium position of the Fermi energy $\mu_l =\mu_r$ lies 
near the first heavy--hole resonance, thus, providing  ferromagnetic order at
zero bias.

Since GaMnAs represents a ternary alloy, disorder from Mn$_{\mathrm{Ga}}$ substitution must be accounted for to capture its electronic structure.
In other ternary alloys, such as AlGaAs or CdZnSe, disorder effects lead to features, such as direct to indirect gap transitions and band bowing \cite{Li1991:PRB,Li1992:PRB}.

In addition, unwanted defects, such as antisites, interstitials, Mn complexes, vacancies, etc, as well as 
additional defects arising from the heterointerfaces must be expected or  have been observed 
\cite{Richardella2010:S,Burch2008:JMMM}.
Clearly, the complexity of the material is such that a realistic predictive electronic structure calculation from first principles is practically impossible.  
Hence we must take a completely different approach and model disorder phenomenologically, taking  doping level and basic 
properties of Mn$_{\mathrm{Ga}}$ 
for guidance. 
Here disorder effects in the GaMnAs layer are modeled by performing a
configurational average over structures with randomly selected
onsite and hopping matrix elements of the tight-binding Hamiltonian
in the Mn doped region.   We choose a 5\% Mn$_{\mathrm{Ga}}$ concentration in the well and model random (uncorrelated) substitutional disorder.
If a Mn ion is present at a given  lattice site in the well the
onsite energy is shifted according to a Gaussian distribution around
a mean onsite energy--shift of 80 meV and a standard deviation of 20
meV, which are reasonable values according to data available for the isolated Mn$_{\mathrm{Ga}}$ acceptor, as well as  recent experimental
results for Ga$_{1-x}$Mn$_x$As for x~2-6\% \cite{Richardella2010:S,Ohya2011:NP}. The hopping matrix element is sampled
according to a Gaussian with 5, 10, and 20\%  standard deviation
($\sigma_t$) of its bulk value $t$. This increase is in hopping matrix variation is used to simulate an increasing level of defect concentration, for fixed Mn$_{\mathrm{Ga}}$ concentration.  
For each such randomly selected  Hamiltonian the transport
problem is solved self--consistently and the I-V curve is computed.  Final results for I-V curve, magnitude of magnetization and spin current polarization, etc.,  are obtained
by averaging over individual results obtained for these configurations.  Typically 300 configurations
are used to perform this average.  In this simulation we keep constant the number of active Mn spins so that the formation and degree of ferromagnetic order at given bias is determined self--consistently from the hole spin polarization in the well region.
\begin{figure}[!t]
\centering
\includegraphics[width=\linewidth]{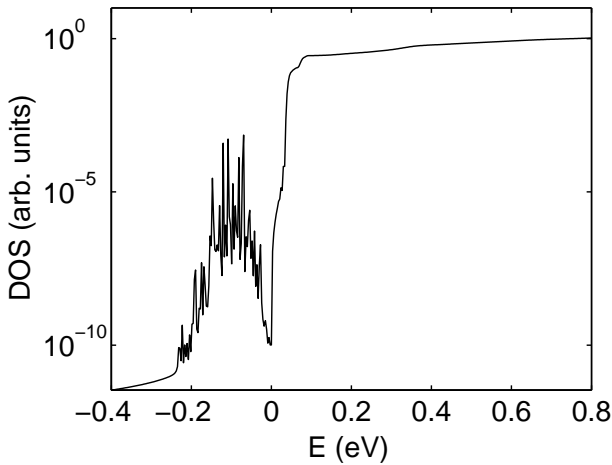}
\caption{Averaged total density of states near the center of the GaMnAs well region for zero bias and
maximum hopping disorder ($\sigma_t = 20\%$). For valence band states ($ E > 0$) integration over in-plane
momentum is taken into account and the result is normalized to the 2D-density of states $D_0 = m^*/(\pi\hbar^2)$.}
\label{fig:dos}
\end{figure}

\begin{figure}[!t]
\centering
\includegraphics[width=\linewidth]{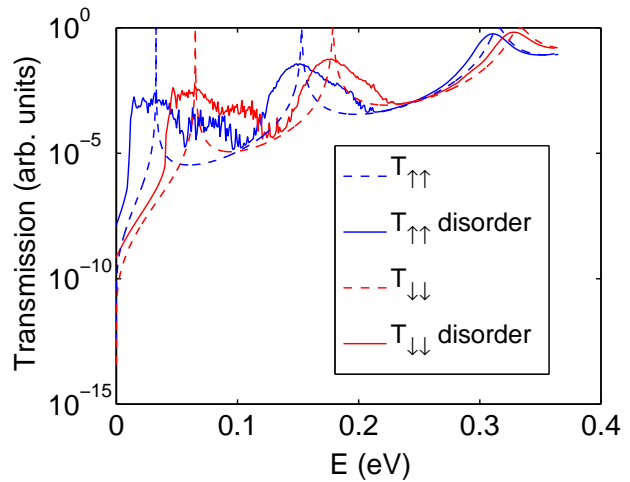}
\caption{(Color online) Spin-dependent transmission probability of the double barrier structure at zero bias with
and without disorder ($\sigma_t = 5 \%$).}
\label{fig:T}
\end{figure}

Figure~\ref{fig:dos} shows the  averaged density of states near the band edge in the Ga$_{1-x}$Mn$_x$As well region for zero bias and maximum degree of hopping disorder modeled here ($\sigma_t = 20\%$).  It is seen that the main effect of our model for substitutional disorder leads to to a broadened  isolated impurity state above the valence band edge, whereby the latter clearly displays the steps characteristic for 2d quantization effects in spite of disorder (note the logarithmic scale!).   The defect levels do not provide a genuine band since they are confined to the Ga$_{1-x}$Mn$_x$As well region.  These states may trap holes and contribute to ionized impurity scattering but are not actively involved in tunneling here since they lie below both quasi Fermi levels at up to moderate bias.    The transmission probability versus energy of the incident holes at
zero bias is displayed in Fig.~\ref{fig:T}.  In absence of disorder, 
the GaMnAs top valence band  structure is modeled as that of GaAs plus a self--consistent exchange and the resonances, indicated as dashed lines in
Fig.~\ref{fig:T}, are spin-split by about 30 meV.   
Taking into account 
disorder, see solid curves for  $\sigma_t =5\%$  in Fig.~\ref{fig:T})  leads to spectral broadening and a shift  of the resonances deeper into the valence band (anti--bonding effect)  of the
resonances.  An increase in overlap of the transmission peaks for spin--up and spin--down holes under disorder is
particularly pronounced for the first heavy--hole resonance since it
is most sensitive to potential fluctuations. 
For low contact temperatures, however, this effective spin splitting ensures spin--dependent tunneling rates even at a level of disorder where the non--monotonic increase of the current with applied bias is practically lost, as shown below.

Contributions from the light--hole band lead to additional resonances.   
Their inclusion would call for a higher--dimensional tight--binding model which properly captures nonparabolicity parallel to the heterointerfaces and goes beyond the scope of this paper.  We just point out that,  for the present structure a
light--hole--band resonance would be expected somewhere between the first two
heavy--hole--associated resonances, contributing  to a further masking of negative differential conductance \cite{Ohya2011:NP}.

\begin{figure}[!t]
\centering
\includegraphics[width=\linewidth]{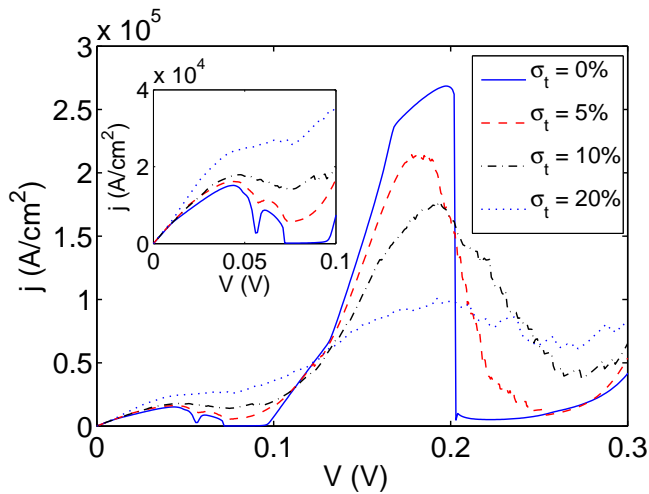}
\caption{(Color online) $IV$-characteristics of a magnetic double barrier structure due to heavy--hole associated bands for
different degrees of disorder, i.e., standard deviations $\sigma_t$ of the hopping matrix elements, as explained in the text.
The inset shows the IV-curve in the voltage range in which the transport takes place via the first heavy-hole subband.}
\label{fig:IVgreen}
\end{figure}

Discontinuous
second derivatives of the IV-curve obtained from tunneling spectroscopy  have been attributed to 
quantization effects in the GaMnAs quantum well.    Regions of negative
differential conductivity, however, have not been observed directly in the IV-curve
\cite{Ohya2011:NP}. This suggests that disorder may play a
considerable role, similar to  amorphous silicon quantum
wells for which weak signatures of resonances have been  predicted and observed
\cite{Miyazaki1987:PRL,Li1993:PRB}. The current-voltage I--V
characteristics for an increasing degree of disorder is
plotted in Fig.~\ref{fig:IVgreen}. Our model reveals that the first
region of negative differential resistance, corresponding to the
first heavy hole resonance, does not disappear until  considerable
hopping disorder of about 10\%, modeled as variance in $t$,  is assumed.
This relatively high value needed to
flattened the IV-curve in our simulations suggests that in real
samples defects other than Mn$_{\mathrm{Ga}}$, such as  interstitials, antisite defects or
voids, may play a considerable role for blurring the resonances.

\begin{figure}[!t]
\centering
\includegraphics[width=\linewidth]{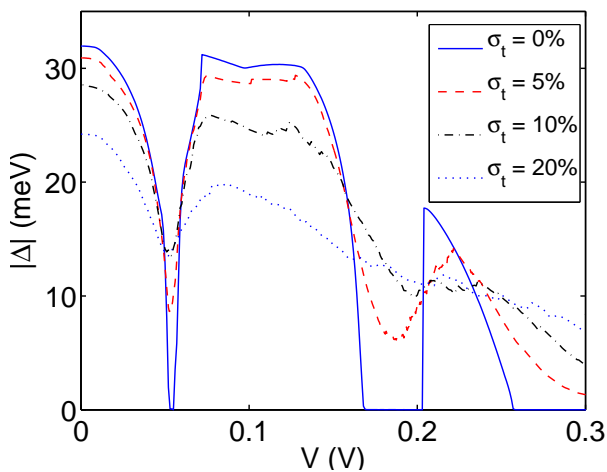}
\caption{(Color online) The configuration averaged spin splitting $|\Delta|$ in the quantum well
 as a function of the applied bias for different degrees of disorder.}
\label{fig:Delta}
\end{figure}

\begin{figure}[!t]
\centering
\includegraphics[width=0.9\linewidth]{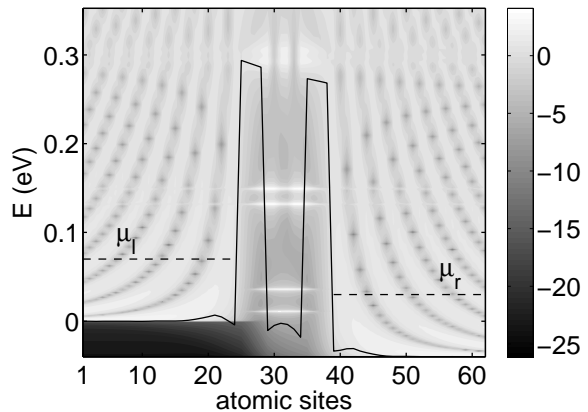}
\caption{Logarithmic local density of states (LDOS)
as a function of energy at the bias $V = 0.04$ V (before the first current maximum). The
self-consistent band profile is indicated by the solid line.}
\label{fig:ldos0}
\end{figure}

\begin{figure}[!t]
\centering
\includegraphics[width=0.9\linewidth]{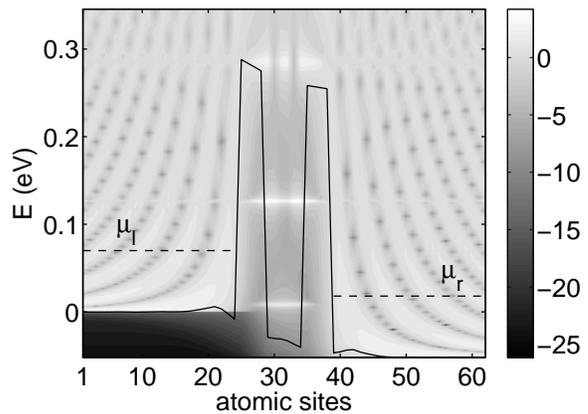}
\caption{Logarithmic local density of states (LDOS)
as a function of energy at the bias $V = 0.052$ V (at the first current peak). The
self-consistent band profile is indicated by the solid line.}
\label{fig:ldos1}
\end{figure}

\begin{figure}[!t]
\centering
\includegraphics[width=0.9\linewidth]{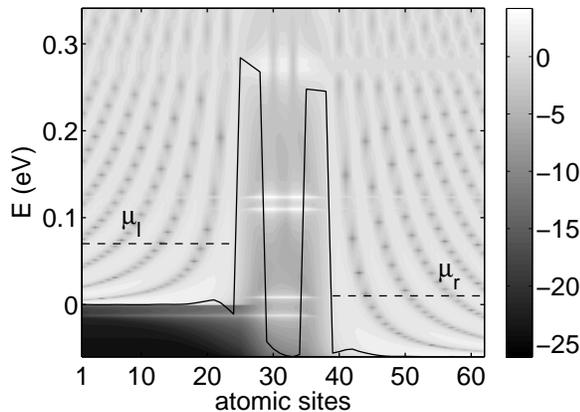}
\caption{Logarithmic local density of states (LDOS)
as a function of energy at the bias $V = 0.06$ V (at off-resonance conditions). The
self-consistent band profile is indicated by the solid line.}
\label{fig:ldos2}
\end{figure}

\begin{figure}[!t]
\centering
\includegraphics[width=\linewidth]{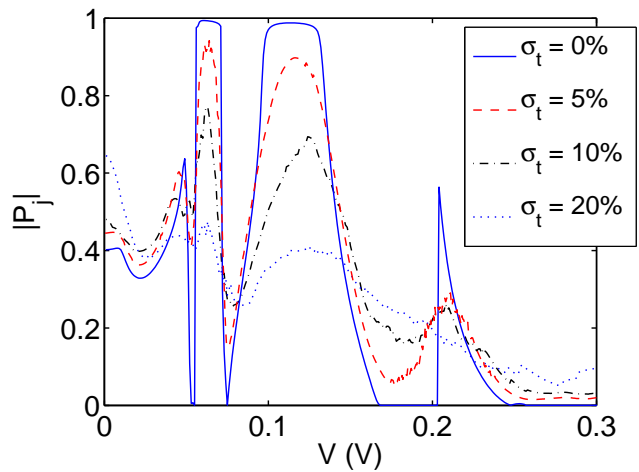}
\caption{(Color online) Averaged current spin polarization $|P_j|$ versus
applied bias $V$ for different degrees of disorder.}
\label{fig:pj}
\end{figure}

In Fig.~\ref{fig:Delta} the  average
exchange band spin splitting $|\Delta|$ in the quantum well is plotted versus applied bias, showing
that the degree of ferromagnetic order  in this structure  can be controlled by the applied bias.  At
the positions of the current maxima (compare to
Fig.~\ref{fig:IVgreen}) the quantum well is
populated by the contacts with unpolarized holes, resulting in a breakdown of the ferromagnetic order.
The degree of  bias control is moderated for increasing defect concentrations.   However, our averaging procedure may  exaggerate this reduction since it  corresponds to 
a physical situation where, in plane, there is a coexistence of (uncorrelated) ferromagnetic and non--ferromagnetic domains.  If on the other hand, in--plane ferromagnetic order is established or destroyed coherently (in correlated fashion),
the degree of bias control is underestimated by our sampling procedure.   This makes us suggest to perform spin--sensitive tunneling spectroscopy, ideally in presence of an external magnetic field,  since the current spin polarization is predicted to be a more sensitive signature to determine the degree of ferromagnetic order in the sample than resonances in the I--V curve.  In fact the latter are predicted to occur at a bias when the sample is in the nonmagnetic state only.  

To illustrate the changes in the magnetic state of the quantum well near  the first current peak at $V = 0.052$~V, we plot the local density of states (LDOS) closely below resonance ($V = 0.04$~V, Fig.~\ref{fig:ldos0}),
 at resonance ($V = 0.052$~V, Fig.~\ref{fig:ldos1}), and above resonance ($V = 0.06$~V, Fig.~\ref{fig:ldos2}).
 At zero bias (not shown in Fig.~\ref{fig:ldos0}) ferromagnetic order is present, placing the Fermi energy between the spin split subband edges.  The spin up level is found at a lower energy than the spin down level. (Note that  we use an inverted energy scheme for the valence band.)  This situation is achieved by matching the contact doping level to the lowest (heavy--hole) resonance in the heterostructure.  
 As bias is increased both spin--up and spin--down subband become accessible from the emitter side, however, only the lower spin--up subband is accessible from the collector side.  
 This maintains hole spin polarization, however, it decreases with increasing bias, leading to a continuous decrease in $\Delta$ shown in Fig.~\ref{fig:Delta}.   
 While hole spin polarization decreases, hole density increases in the well region partially screening the applied bias.  As the applied bias is increased resonance is reached.   Hole charge and exchange splitting in the well 
 are no longer sufficient to keep the spin--down subband above the collector quasi--Fermi level and ferromagnetic order collapses, as shown in  Fig.~\ref{fig:ldos1}.  
Both  subbands are equally flooded from both contacts leading to an unpolarized hole gas in the well and, hence,
to a destruction of ferromagnetic order.   Note the change in the effective potential profile in the well from convex up in Fig. ~\ref{fig:ldos0} to almost linear in  Fig.~\ref{fig:ldos1}.   If the bias is increased further the levels are pushed closely below the 
emitter valence band edge and resonance to the emitter is suppressed.  However, any  small perturbation triggers the system into a ferromagnetic state, in which the spin up level preferrentially 
is filled from the collector side (see Fig.~\ref{fig:ldos2}). Thus,  inspection of the LDOS-plots shows
that the ferromagnetic state in the quantum well is determined by the relative position of the well subband edges and the contact
quasi--Fermi levels. An appropriate tailoring of these levels allows to change the hole gas polarization under bias and thereby provides electrical
control of the ferromagnetic state in the well. In our case we consider only a two--terminal configuration with source and drain contacts but the  use of additional gates in transverse direction
(multi--terminal configurations)  provides an additional control knob to move the subbands.
Although these structures are very difficult to realize in practice, they have been studied in a
recent experiment \cite{Ohya2010:APL}.

The change of the magnetic state in the well is directly reflected  
in the (collector) current spin polarization, as shown in
Fig.~\ref{fig:pj}.  If the entire quantum well becomes nonmagnetic, the spin
density in the well vanishes and the collector current becomes
unpolarized.  A polarized current indicates exchange--split subbands in the well, demonstrating ferromagnetism in the well. The experimental probing of the current spin
polarization at the collector side therefore would give important additional
information to confirm the interpretation of recent experiments regarding
size quantization effects in GaMnAs quantum wells
\cite{Ohya2011:NP}.

\section{Conclusions and Outlook}\label{sec:sum}

In summary, we have used a steady--state transport model
to investigate the role of structural disorder on the interplay of ferromagnetic order and resonant tunneling in
double barrier structures with a GaMnAs quantum well. Ferromagnetic exchange, as
well as the hole Coulomb interaction are treated within
a self--consistent mean--field approximation. Disorder effects are modeled  by random variation  of onsite and hopping matrix elements of the
tight-binding Hamiltonian according to basic experimental findings on the Mn$_{\mathrm{Ga}}$ acceptor.    In this work we have modeled an electronic structure with an isolated impurity band, as supported by most of the recent experiments.  

For samples which, at zero bias, exhibit  ferromagnetic order in the GaAsMn well we predict that  ferromagnetic
order is destroyed under bias near (the first
heavy-hole) resonance.  While at resonance 
the well region is flooded by holes these are overall  {\it unpolarized} thus prohibiting  the communication of ferromagnetic ordering amongst Mn$_{\mathrm{Ga}}$ sites in the well.   
Within our model we thus are able to provide a possible explanation for the
absence of exchange splitting near resonances, as observed in recent
tunneling spectroscopy measurements on thin GaMnAs
layers \cite{Ohya2011:NP}. Furthermore we find that an experimental investigation of the spin
polarization of the collector current can give information about the
presence of magnetism in the quantum well. Such a measurement is more revealing regarding the ferromagnetic state than the search for resonances in the I--V curve, since the former is more robust against disorder than the latter.  We find that disorder tends to suppress  negative differential resistance
region in the IV-curve.   Although our model is merely qualitative it indicates that substitution disorder from Mn doping alone is not sufficient to explain the absence negative differential conductivity in experiment.  

There are a number of open question which should be addressed in the future.  One pertains to the correlation length of ferromagnetic order parallel to the heterointerface.  It has direct influence on the spin--valve action and its bias--control of  ferromagnetic heterostructures.   Furthermore, one may ask to what extent magneto--transport  experiments can distinguish between the two main electronic structure models (with and without isolated impurity band) discussed in the literature.

\section{Acknowledgment}

This work has been supported by the FWF project P21289-N16.


\end{document}